\documentclass[12pt]{elsarticle}

\usepackage{graphicx}

\begin{document}

\begin{frontmatter}

\title{Analytical expression for the convolution of a Fano line profile with a Gaussian}

\author{S.~Schippers}
\ead{schippers@jlug.de}
\ead[url]{http://www.uni-giessen.de/amp}
\address{I. Physikalisches Institut, Justus-Liebig-Universit\"{a}t Gie{\ss}en, Heinrich-Buff-Ring 16, 35392 Giessen, Germany}

\begin{abstract}
Asymmetric Fano line profiles are frequently encountered, e.\,g., in the photoionization spectra of atoms and ions. For
the fitting of spectral line profiles to experimental spectra the line profiles have to be convolved with the
experimental window function. The latter is often taken to be a Gaussian. It is shown that the  convolution can be represented by a rather simple analytic expression involving the Faddeeva function for the evaluation of which efficient and accurate numerical algorithms are available.
\end{abstract}

\end{frontmatter}

\section{Introduction}

Asymmetric line profiles are frequently encountered, e.g.\ in atomic photoionization \cite{Schulz1996a,Kjeldsen2006a,Schippers2016} or photorecombination \cite{Knapp1995a,Schippers2002a,GonzalezMartinez2005a}, due to a quantum mechanical interference between resonant and a nonresonant ionization or recombination  pathways. According to the quantum mechanical analysis of Fano \cite{Fano1961} the photoionization cross section  as a function of photon energy $E$ in the vicinity of the resonance energy $E_\mathrm{res}$ can be represented as \cite{Fano1965}
\begin{equation}\label{eq:fano0}
\sigma(\epsilon) = \sigma_0 + \sigma_1\frac{(q+\epsilon)^2}{1+\epsilon^2}
\end{equation}
where $q$ is the asymmetry parameter and $\sigma_0$ and $\sigma_1$ are slowly varying functions of the reduced energy $\epsilon = 2(E-E_\mathrm{res})/\Delta_L$. Here, $\Delta_L$ is the natural (Lorentzian) line width.

For the extraction of the resonance parameters $\sigma_1$, $E_\mathrm{res}$, $\Delta_L$ and $q$ from the experimental data, the measured resonance lines can be fitted by a Fano profile. In such fits the experimental photon energy distribution (window function) has to be taken into account. In many cases the experimental window function can be represented as a Gaussian where the Gaussian full width at half maximum (FWHM) corresponds to the experimental energy spread. Thus, for a fit to the experimental data the Fano profile has to be convolved with a Gaussian.

The situation is similar to emission spectroscopy of hot gases where Doppler broadening results in Voigt line profiles, i.e., the convolution of a Lorentzian with a Gaussian. It is well known (see, e.g., \cite{Humlicek1979,Wells1999a,Schreier2011,Zaghloul2011}) that the Voigt profile can be calculated efficiently from the Faddeeva function \cite{Abramowitz1964, cerf} (a scaled complex error function).
Here, it is shown that also the convolution of a Fano profile with a Gaussian can be represented by the Faddeeva function.   The resulting formula allows for a fast and accurate evaluation of the convolution, e.g., in peak fitting routines. It is mentioned that a different, more complex formula has been published earlier without its derivation \cite{Fang1998}. It seems, that this formula has not received much attention since even in more recent work the convolution of a Fano profile with a Gaussian has only been treated approximately or by numerical integration \cite{Teodorescu1997,Liu2003}.

The present paper, which refines an earlier preprint \cite{Schippers2012c}, is organized as follows. In section \ref{sec:voigt} the calculation of the Voigt profile from the Faddeeva function is reviewed. In section \ref{sec:cerf} some relevant properties of the Faddeeva function are presented. An expression for the convolution of the Fano profile with a Gaussian in terms of the Faddeeva function is derived in section~\ref{sec:fanogauss}. A conclusive summary is given in section \ref{sec:summary}.

\section{The Voigt profile}\label{sec:voigt}

The convolution of a Lorentzian line profile
\begin{equation}\label{eq:lor}
{L}(E) = A\frac{2}{\pi}\frac{\Delta_L}{4(E-E_{\rm res})^2+\Delta_L^2}
\end{equation}
with a Gaussian
\begin{equation}\label{eq:gau}
{G}(E) = \frac{2}{\Delta_G}\sqrt{\frac{\ln 2}{\pi}}\exp\left[-\frac{4(\ln2)\,E^2}{\Delta_G^2}\right],
\end{equation}
yields the Voigt profile
\begin{eqnarray}
{V}(E) &=& \int_{-\infty}^\infty {L}(E')\,{G}(E'-E)\,dE'\\
     &=& A\frac{4\sqrt{\ln 2}}{\pi^{3/2}\Delta_G}\int_{-\infty}^\infty
     \frac{\Delta_L}{4(E'-E_{\rm res})^2+\Delta_L^2} \exp\left[-\frac{4(\ln2)\,(E-E')^2}{\Delta_G^2}\right]
     dE'.\nonumber\label{eq:voi}
\end{eqnarray}
The profiles in Eqs.~\ref{eq:lor} and \ref{eq:gau} are normalized such that
\begin{equation}\label{eq:norm}
\int {L}(E)\, dE = A \mbox{\rm ~and~} \int {G}(E)\, dE = 1.
\end{equation}
The widths $\Delta_L$ and $\Delta_G$ are the Lorentzian and Gaussian FWHM,
respectively. With the definitions
\begin{equation}\label{eq:tdef}
t = \frac{2\sqrt{\ln 2}(E'-E)}{\Delta_G},\;\;
x = \frac{2\sqrt{\ln 2}(E_{\rm res}-E)}{\Delta_G},\;\;\textrm{and}\;
y = \frac{\Delta_L\sqrt{\ln 2}}{\Delta_G},
\end{equation}
Eq.~\ref{eq:voi} transforms into
\begin{equation}\label{eq:voigt}
{V}(E) = A\frac{2\sqrt{\ln 2}}{\Delta_G\sqrt{\pi}}\,\frac{1}{\pi}\int_{-\infty}^\infty \frac{y
e^{-t^2}}{(t-x)^2+y^2}dt= A\frac{2\sqrt{\ln 2}}{\Delta_G\sqrt{\pi}}\,\Re[w(z)]
\end{equation}
where $w(z)$ denotes the Faddeeva function  and $z = x+iy$.

\section{Some properties of the Faddeeva function}\label{sec:cerf}

\begin{figure}
\centering{\includegraphics[width=\textwidth]{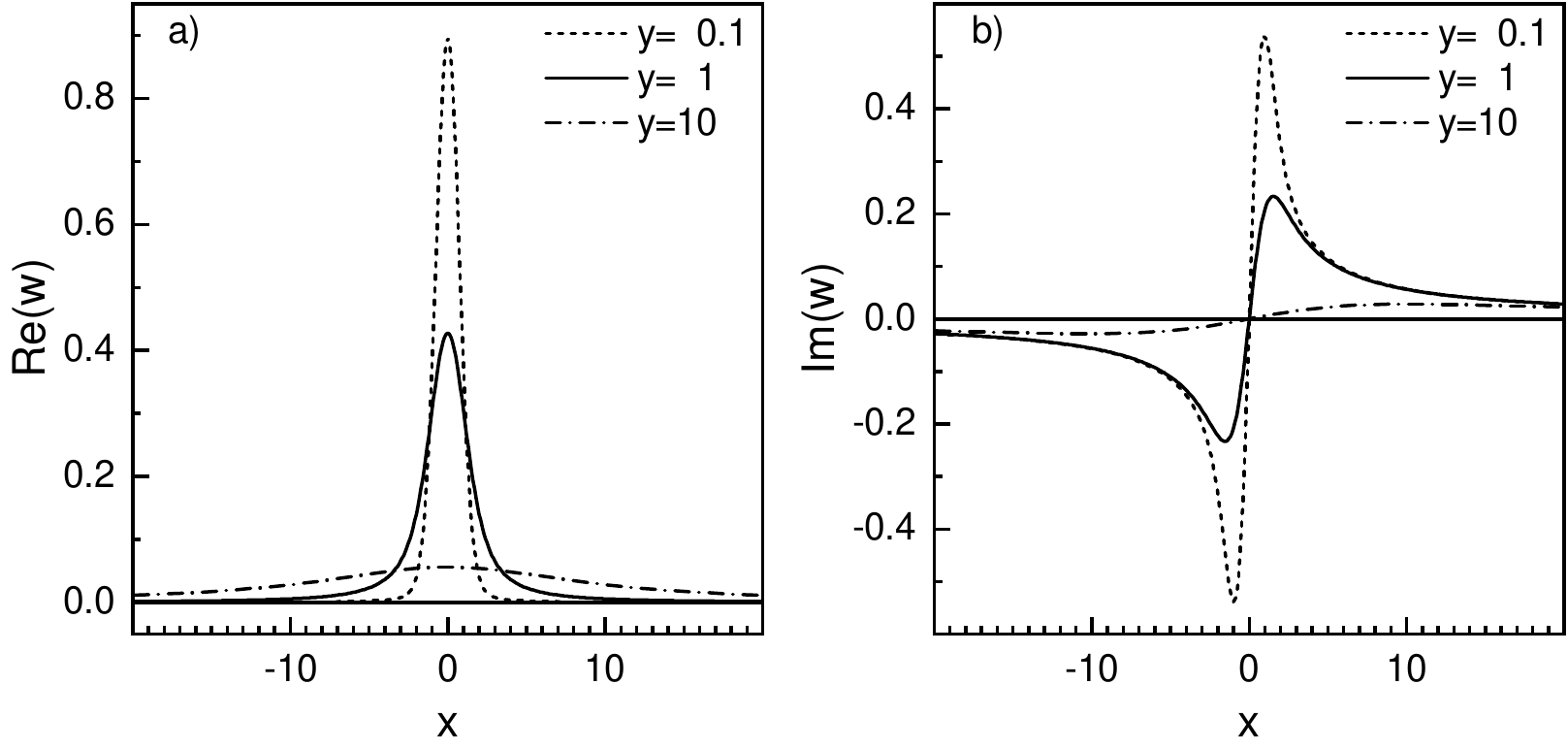}}
\caption{\label{fig:w}Real part (panel a) and imaginary part (panel b) of the Faddeeva function $w(x+iy)$ as function of the scaled energy $x = 2\sqrt{\ln 2}(E_{\rm res}-E)/\Delta_G$ for different ratios $y = \Delta_L\sqrt{\ln 2}/\Delta_G$ of Lorentzian and Gaussian widths. The real part is essentially the Voigt profile (Eq.~\ref{eq:voigt}). The imaginary part occurs in addition in the convolution of a Fano profile with a Gaussian (Eq.~\ref{eq:c3}).}
\end{figure}

The Faddeeva function is a scaled complex error function. It is defined for $\Im z = y >0$ as \cite{Abramowitz1964,cerf,Poppe1990,Sampoorna2007}
\begin{equation}\label{eq:w}
w(z)= e^{-z^2}\textrm{erfc}(-iz) = \frac{i}{\pi}\int_{-\infty}^\infty \frac{e^{-t^2}}{z-t}dt.
\end{equation}
Its real and imaginary parts are
\begin{equation}\label{eq:rew}
\Re[w(z)] = \frac{1}{\pi}\int_{-\infty}^\infty \frac{y e^{-t^2}}{(t-x)^2+y^2}dt
\end{equation}
and
\begin{equation}\label{eq:imw}
\Im[w(z)]=\frac{-1}{\pi}\int_{-\infty}^\infty \frac{(t-x)e^{-t^2}}{(t-x)^2+y^2}dt.
\end{equation}
Figure \ref{fig:w} displays $\Re[w(x+iy)]$ and $\Im[w(x+iy)]$  as functions of the scaled energy $x$  for different ratios $y$ of Lorentzian and Gaussian widths.

For later use we now calculate the integral
\begin{equation}\label{eq:intt2}
  {I}_2(x,y) = \frac{1}{\pi}\int_{-\infty}^\infty \frac{t^2 e^{-t^2}}{(t-x)^2+y^2}dt.
\end{equation}
To this end we define (cf.\ Eq.~\ref{eq:w})
\begin{equation}\label{eq:weta}
w_\eta(z)= \frac{i}{\pi}\int_{-\infty}^\infty \frac{e^{-\eta t^2}}{z-t}dt.
\end{equation}
With this definition
\begin{equation}\label{eq:intt2a}
{I}_2(x,y) = \frac{-1}{y}\,\Re\left[\frac{d w_\eta(z)}{d\eta}\right]_{\eta=1}.
\end{equation}
For the calculation of the derivative on the right-hand side of this equation we exploit the identity
$w_\eta(z)= w(\sqrt{\eta}z)$ which follows from the substitution $(\sqrt{\eta}t) \to t$ in
Eq.~\ref{eq:weta}. This yields
\begin{eqnarray}
  \frac{d w_\eta(z)}{d\eta} &=& \frac{d w(\sqrt{\eta}z)}{d\eta} =
  \frac{d(\sqrt{\eta}z)}{d\eta}\frac{d w(\sqrt{\eta}z)}{d (\sqrt{\eta}z)}\nonumber\\ & = &
  \frac{z}{2\sqrt{\eta}}\left(-2(\sqrt{\eta}z) w(\sqrt{\eta}z)+ \frac{2i}{\sqrt{\pi}}\right)\label{eq:dw1}
\end{eqnarray}
where in the last step we have used the identity \cite{Abramowitz1964,cerfderiv}
\begin{equation}\label{eq:dw2}
  \frac{dw(z)}{dz} = -2zw(z)+\frac{2i}{\sqrt{\pi}}.
\end{equation}
Combining Eqs.~\ref{eq:intt2a} and \ref{eq:dw1} yields
\begin{equation}
{I}_2(x,y) = \frac{1}{y}\,\Re\left[z^2w(z)-i\frac{z}{\sqrt{\pi}}\right] = \left(\frac{x^2}{y}-y\right)\Re(w)-2x\,\Im(w) +\frac{1}{\sqrt{\pi}}.\label{eq:intt2b}
\end{equation}

From equation \ref{eq:dw2} and by exploiting the Cauchy-Riemann relations, the partial derivatives of $w(z)$ with respect to $x$ and $y$ are easily calculated as
\begin{eqnarray}
\frac{\partial\Re(w)}{\partial x} &=& \;\;\;\frac{\partial\Im(w)}{\partial y} = -2x\Re(w)+2y\Im(w),\\
\frac{\partial\Im(w)}{\partial x} &=& -\frac{\partial\Re(w)}{\partial y} =
-2y\Re(w)-2x\Im(w)+\frac{2}{\sqrt{\pi}}.
\end{eqnarray}
These are useful, e.g., in least-squares fitting routines.

\section{Convolution of a Fano profile with a gaussian}\label{sec:fanogauss}

For the purpose of peak fitting we define the Fano line profile somewhat differently as suggested by Eq.~\ref{eq:fano0}, i.e.
\begin{equation}\label{eq:fano1}
  {F}(E) = \frac{2a}{q^2\Delta_L{\pi}}\left[\frac{(q+\epsilon)^2}{1+\epsilon^2}-1\right]
\end{equation}
with a positive amplitude $a>0$. For $q\to \infty$ the Fano profile as defined by Eq.~\ref{eq:fano1} approaches a symmetric Lorentzian (Eq.~\ref{eq:lor}), i.\,e.\ ${F}(E) \to {L}(E)$. Moreover, this definition of the Fano profile is consistent with the theoretical treatment of the atomic photoabsorption cross section \cite{Rost1997}.

With $t$, $x$ and $y$ from Eq.~\ref{eq:tdef} the convolution with a Gaussian as defined by Eq.~\ref{eq:gau} can be expressed as

\begin{eqnarray}
  {C}(E) &=& \int_{-\infty}^\infty  {F}(E')\, {G}(E'-E)\,dE'\\ &=& \frac{2a}{q^2\Delta_L{\pi}}\,\left[\frac{1}{\sqrt{\pi}}\int_{-\infty}^\infty
   \frac{[qy+(t-x)]^2e^{-{t^2}}}{(t-x)^2+y^2}dt-1\right]\label{eq:c1}\\
   &=& \frac{2a}{q^2\Delta_L\sqrt{\pi}}\times \nonumber\\ &&\left[\frac{1}{\pi}\int_{-\infty}^\infty
   \frac{[q^2y^2+2qy(t-x)+t^2-2x(t-x)-x^2]e^{-{t^2}}}{(t-x)^2+y^2}dt-\frac{1}{\sqrt{\pi}}\right]\nonumber\\
   &=& \frac{2a}{q^2\Delta_L\sqrt{\pi}}\times \nonumber\\ &&\left\{q^2y\Re(w)-2qy\Im(w)+{I}_2+2x\Im(w) -\frac{x^2}{y}\Re(w)-\frac{1}{\sqrt{\pi}}\right\}\nonumber\\
   &=&\frac{2ay}{q^2\Delta_L\sqrt{\pi}}\left\{q^2\Re(w)-2q\Im(w)+\frac{x^2}{y^2}\Re(w) -\Re(w)-\right.\nonumber\\ & & \hspace*{3cm}\left.
   \frac{2x}{y}\Im(w)+\frac{1}{y\sqrt{\pi}}+\frac{2x}{y}\Im(w)-\frac{x^2}{y^2}\Re(w)-\frac{1}{y\sqrt{\pi}}\right\}\nonumber\\
   &=&\frac{a}{q^2} \frac{2\sqrt{\ln 2}}{\Delta_G\sqrt{\pi}}\biggl\{(q^2-1)\Re(w)-2q\Im(w)\biggr\}\label{eq:c2}.
\end{eqnarray}
In this derivation the definitions of $\Re(w)=\Re[w(z)]$ (Eq.~\ref{eq:rew}), $\Im(w)=\Im[w(z)]$ (Eq.~\ref{eq:imw}), and ${I}_2$ (Eqs.~\ref{eq:intt2} and \ref{eq:intt2b}) have been used.
It is easily seen that  ${C}(E)\to {V}(E)$ for $q\to \infty$ as expected and that the asymmetry of the profile ${C}(E)$ is due to the additional term containing the imaginary part $\Im[w(z)]$ (Fig~\ref{fig:w}b) of the Faddeeva function $w(z)$. Since $\Im[w(z)]$ (Eq.~\ref{eq:imw}) is an odd function of the scaled energy $x$  it does not contribute to the integrated line strength, i.e.
\begin{equation}
\int_{-\infty}^\infty{C}(E)\,dE =  \int_{-\infty}^\infty{V}(E)\,dE
\end{equation}
holds. From the comparison of the real part of $C(E)$ (Eq.~\ref{eq:c2}) with the Voigt profile $V(E)$ (Eq.~\ref{eq:voigt}) one obtains the following relation between the  peak area $A$ and the amplitude $a>0$:
\begin{equation}\label{eq:Aa}
A = a\frac{q^2-1}{q^2}
\end{equation}
For $\vert q\vert < 1$ the peak area becomes negative. This indicates that the resonance profile produces a dip (window resonance) in the absorption cross section  \cite{Rost1997}. Inserting  Eq.~\ref{eq:Aa} in Eqs.~\ref{eq:fano1} and \ref{eq:c2} finally yields
\begin{equation}\label{eq:fano2}
  {F}(E) = \left\vert\frac{A}{q^2-1}\right\vert\frac{2}{\Delta_L{\pi}}\left[\frac{(q+\epsilon)^2}{1+\epsilon^2}-1\right]
\end{equation}
and
\begin{equation}\label{eq:c3}
  {C}(E) = \left\vert\frac{A}{q^2-1}\right\vert \frac{2\sqrt{\ln 2}}{\Delta_G\sqrt{\pi}}\left\{(q^2-1)\Re[w(z)]-2q\Im[w(z)]\right\}.
\end{equation}

\section{Summary}\label{sec:summary}

It has been demonstrated that the convolution of a Fano line profile (Eq.~\ref{eq:fano2}) with a Gaussian (Eq.~\ref{eq:gau}) can be evaluated analytically (Eq.~\ref{eq:c3}) by using the Faddeeva function (scaled complex error function) $w(z)$ with $z = x+iy$ and with $x$ and $y$ from Eq.~\ref{eq:tdef}.  Various fast and accurate algorithms for computing the Faddeeva function have been described in the literature \cite{Humlicek1979,Wells1999a,Zaghloul2011,Gautschi1969,Gautschi1970,Humlicek1982,Poppe1990a,Letchworth2007,Abrarov2009}. Their performances have been critically evaluated repeatedly \cite{Schreier2011,Zaghloul2011,Schreier1992,Schreier2008}.  According to the findings of Zaghloul and Ali \cite{Zaghloul2011} their algorithm is the most accurate to date. An implementation of this algorithm in the Fortran programming language has been published  \cite{Zaghloul2016} and a modified version of this algorithm coded in the programming language C$^{++}$ is available from the internet \cite{Johnson2012}.

The line profile $C(E)$ (Eq.~\ref{eq:c3}) has been implemented by the author as a user-supplied fit function for the commercial software \textsc{Origin} \cite{Origin}. The implementation is available from the author upon request. It has already been successfully used in various contexts \citep[e.g.][]{Schippers2002a,Schippers2003b,Scully2006a}. It should be noted that the here introduced derivations can also be applied to related line shapes. An example has been provided recently in the context of precision spectroscopy of atomic hydrogen \cite{Beyer2017}.

\section*{Acknowledgments}
The author would like to thank Alfred M\"uller and Thomas Udem for helpful discussions.




\end{document}